\documentclass[12pt,preprint]{aastex}

\shorttitle{Spectral models for \ion{Fe}{3}}
\shortauthors{Bautista et~al.}
\begin{document}
\title{ATOMIC DATA AND SPECTRAL MODEL FOR \ion{Fe}{3}} 

\author{Manuel A. Bautista} 
\affil{Department of Physics, Western Michigan University,
Kalamazoo, MI 49008, USA}
\email{manuel.bautista@wmich.edu}           

\author{Connor P. Ballance}
\affil{Department of Physics, Auburn University, Auburn AL 36849}              

\author{Pascal Quinet} 
\affil{Astrophysique et Spectroscopie, Universit\'e de Mons - UMONS,
B-7000 Mons, Belgium}  
\affil{and IPNAS, B15 Sart Tilman, Universit\'e de Li\`ege, Belgium}

\begin{abstract}
We present new atomic data (radiative transitions rates 
and collision strengths) from large scale calculations and a non-LTE 
spectral model for \ion{Fe}{3}. 
This model is in very good agreement 
with observed astronomical emission spectra, in contrast with previous 
models that yield large discrepancies with observations. 
The present atomic computations employ a combination of atomic physics 
methods, e.g. relativistic Hatree-Fock, the Thomas-Fermi-Dirac potential, and 
Dirac-Fock computation of A-values and R-matrix with intermediate 
coupling frame transformation and Dirac R-matrix. We study the advantages and 
shortcomings of each method. It is found that the Dirac R-matrix collision strengths yield excellent agreement with observations, much improved over previously available models. 
By contrast, the transformation of LS-coupling R-matrix fails to
yield accurate effective collision strengths at around 10$^4$~K, despite using very large configuration expansions, 
due to the limited treatment of spin-orbit effects in the near threshold resonances of the collision strengths.
The present work demonstrates that accurate atomic data for low ionization
iron-peak species is now within reach. 
\end{abstract}


\normalsize

\section{INTRODUCTION}

Reliable non-local thermodynamic equilibrium (NLTE)  modeling of opacities and spectra of low ionization
stages of Fe and iron peak species is of 
paramount importance for 21st Century Astrophysics owing to various 
fundamental research areas that depend on these models.
Such models require accurate transition rates (A-values) and
effective collision strengths for electron impact excitation.
These data sets have been difficult to obtain theoretically despite extensive
efforts over many years (e.g. Bautista and Pradhan 1998 and references therein),while 
experimental determinations remain very sparse or
inexistent. 

The \ion{Fe}{3} spectrum is prominent in various galactic sources such as H~II 
regions, planetary nebulae, and Herbig-Haro objects (e.g. Mesa-Delgado et al. 
2009), stars like $\eta$ Carinae (Johansson et al. 2000), and extragalactic 
objects such as Active Galactic Nuclei (Laor et al. 1997, Vertergaard and 
Wilkes 2001). 
These spectra are potentially useful in diagnosing the physical conditions of the plasmas and determining the gas phase Fe component. However, there are still shortcomings in our ability to interpret the observed spectra. 
Most quantitative analysis of \ion{Fe}{3} spectra reported so far have 
relied on the electron impact excitation rates of \cite{zhang96}
and A-values for forbidden transitions of either
\cite{naharpradhan} or \cite{quinet96}.
While the collisional data accounted for all levels
of the 3d$^6$, 3d$^5$4s, and 3d$^5$4p configurations, the available radiative data for dipole forbidden transitions 
comprise only the 3d$^6$ levels (see 
Bautista and Pradhan 1998), leaving out important 
transitions among 3d$^6$ and 
3d$^5$4s levels. Moreover, these transitions are needed in modeling of fluorescent excitation mechanisms \citep{johansson}.  

The accuracy of current models for \ion{Fe}{3} is questionable. When these are 
applied to analyze spectra of the Orion nebulae, the best studied nebula H~II 
region, they yield great scatter  among line ratio diagnostics and derived 
abundances from different lines. Furthermore, when these data are used in deriving
the Fe abundance from [\ion{Fe}{3}] lines this 
differs by a factor of $\sim$4 from abundances derived from
[\ion{Fe}{4}] lines \citep{rodriguezrubin}. 
Moreover, recent calculations of LS-coupling collision strengths by \cite{mclaughlin02,mclaughlin07} yield significant differences with respect to
those of \cite{zhang96}. Unfortunately, McLaughlin et al. provide no fine structure 
data that can be used in spectral models. 

We have carried out extensive calculations using several independent
atomic structure codes. 
This multi-platform approach has proven successful in 
our previous studies of the K shells of Fe, O, Ne, Mg, Si, S, Ar, Ca, and Ni (e.g. 
Bautista et al. 2003, Garcia et al. 2005, Palmeri et al. 2003a, 2003b). This allows for consistency checks and 
intercomparison. For calculations of atomic structure and A-values we use the
semi-relativistic Hartree--Fock (HFR) code of \cite{cowan},
the multi configuration Dirac-Fock method in the code GRASP \citep{Dyall89, Parpia96}, and
the central Thomas-Fermi-Dirac-Amaldi potential in
AUTOSTRUCTURE \citep{bad86,bad97}. 
HFR provides self consistent optimization for the ground and lowest 
configurations, and allows for semi-empirical corrections of the radial integral. The disadvantage, though, is that for complex systems like the present one 
the method converges only with a small number of configurations. 
GRASP is a fully relativistic code that can yield highly optimized 
wavefunctions, but for complex near neutral systems it only converges 
only for small numbers of configurations. AUTOSTRUCTURE is very efficient in 
treating large configuration expansions and CI effects, but uses more approximate orbitals.

The R-matrix scattering calculations are carried out within two different coupling 
schemes. The first method (RM+ICFT hereafter) consists on the intermediate coupling frame transformation method (ICFT; Griffin et al. 1998)
built upon the traditional LS R-matrix package (Berrington et al. 1995).    
It allows for the computationally intensive inner region calculation to be
carried out in LS coupling 
with the inclusion mass-velocity and Darwin terms, and level-to-level collision 
strengths only being acquired from the transformation of term-resolved K/S matrices
in the outer region. Arguably, the capability of the RM+ICFT method to include 
large CI expansions in description the target, and also large numbers 
of correlation terms in the scattering wavefunction, has the potential to 
yield accurate results. However, while ICFT 
properly transforms the background cross
sections and resonances with large principal quantum numbers, 
the positions of low $n$ resonances that lie within the R-matrix box are
not corrected. Moreover, the large imbalance in the number 
the terms included in close coupling expansion and those used 
in the target description also has the potential lead to spurious resonance structure.

Secondly, we employed a modified version 
     of the Dirac-Coulomb R-matrix scattering 
     package DARC 
(Ait-Tahar et al. 1996) and carried out the calculation in JJ coupling in both the inner 
     and outer regions. The availability of modern parallel supercomputers has removed 
     many of the reservations of carrying out large scale R-matrix calculations with DARC,
 e.g. Ballance (2009).

The rest of this letter is structured as follows: In section 2 we describe the calculations
of transition rates for dipole forbidden transitions.
In section 3 we present our calculations of collision strengths. In section 4 we build a spectral model for the ion and compare its predictions for relative
line intensities with observation of the well known Orion nebula. 
Finally, discussion and conclusions are given in section 5.

\section{Radiative Calculations}

We compute radiative rates (A-values) for dipole forbidden transition among the levels of 3d$^6$ and 3d$^5$4s configurations using the codes AUTOSTRUCTURE and HFR.

AUTOSTRUCTURE is an atomic structure package written by \citet{bad86,bad97} 
initially based  
       upon an earlier
structure program
SUPERSTRUCTURE \citep{eis74}, which computes fine-structure
level energies and radiative rates in a Breit--Pauli
relativistic framework. Single electron orbitals are
constructed by diagonalizing the non-relativistic Hamiltonian,
based on the Thomas--Fermi--Dirac--Amaldi (TFDA) model
potential \citep{eis69}.  
Scaling parameters for the potential are optimized variationally by minimizing 
$LS$ term energies. 

We performed calculations with various different
configuration expansions, starting with
models similar to those of previously published
work and then evolving to 
larger expansions.
Our final expansion includes 36 configurations, based on the 16-configuration 
expansion of 
\cite{zhang96}, but adding pseudo-orbitals 4f, 5s, 5p, and
 5d and various single and 
double promotions out of the 3s and 3p orbitals. A full description of the 
target will be given in a future publication.
The orbitals were optimized 
minimizing the energies of the lowest 40 terms of the 3d$^6$ and 3d$^5$4s, and 
3d$^5$4p configurations.
Fine tuning of the wavefunctions was performed by means of term energy corrections. 
When comparing the predicted energy levels with experimental values (Ekberg 1993)
the agreement is typically within $\sim$10\% for the lowest terms up to about 0.55~Ry and better for higher terms.


    HFR uses the suite of codes of Cowan (1981) within the framework of the 
semi-relativistic Hartree-Fock (HFR) method. 
CI was retained among the following 12 configurations: 3d$^6$, 3d$^5$4s, 
3d$^5$5s, 3d$^5$4d, 3d$^4$4s$^2$, 3d$^4$4p$^2$, 3d$^4$4d$^2$, 3d$^4$4s4d, 
3d$^4$4s5s, 3s3p$^6$3d$^7$, 3s3p$^6$3d4s, 3s3p$^6$3d$^5$4s$^2$. This configuration list extends considerably that of Quinet (1996).  In addition to the explicit introduction of CI, the interactions with more distant configurations were simulated through semi-empirical adjustment of the Slater parameters and the inclusion of additional effective parameters, such as $\alpha$ and $\beta$, associated with the excitations out of the 3s and 3p subshells into the 3d (Trees 1951a,b; Racah 1952). The fitting procedure was applied to the 3d$^6$ and 3d$^5$4s configurations with the experimental energy levels published by Ekberg (1993). The ab initio HFR values for the Slater parameters within configurations other than 
those included in the fitting procedure and for the CI integrals, R$^k$, were scaled down by a factor 0.90 as recommended by Cowan (1981) while the ab initio values of all the spin-orbit integrals, computed by the Blume-Watson method, were used without scaling.

\begin{figure}
\resizebox{\hsize}{14cm}{\includegraphics{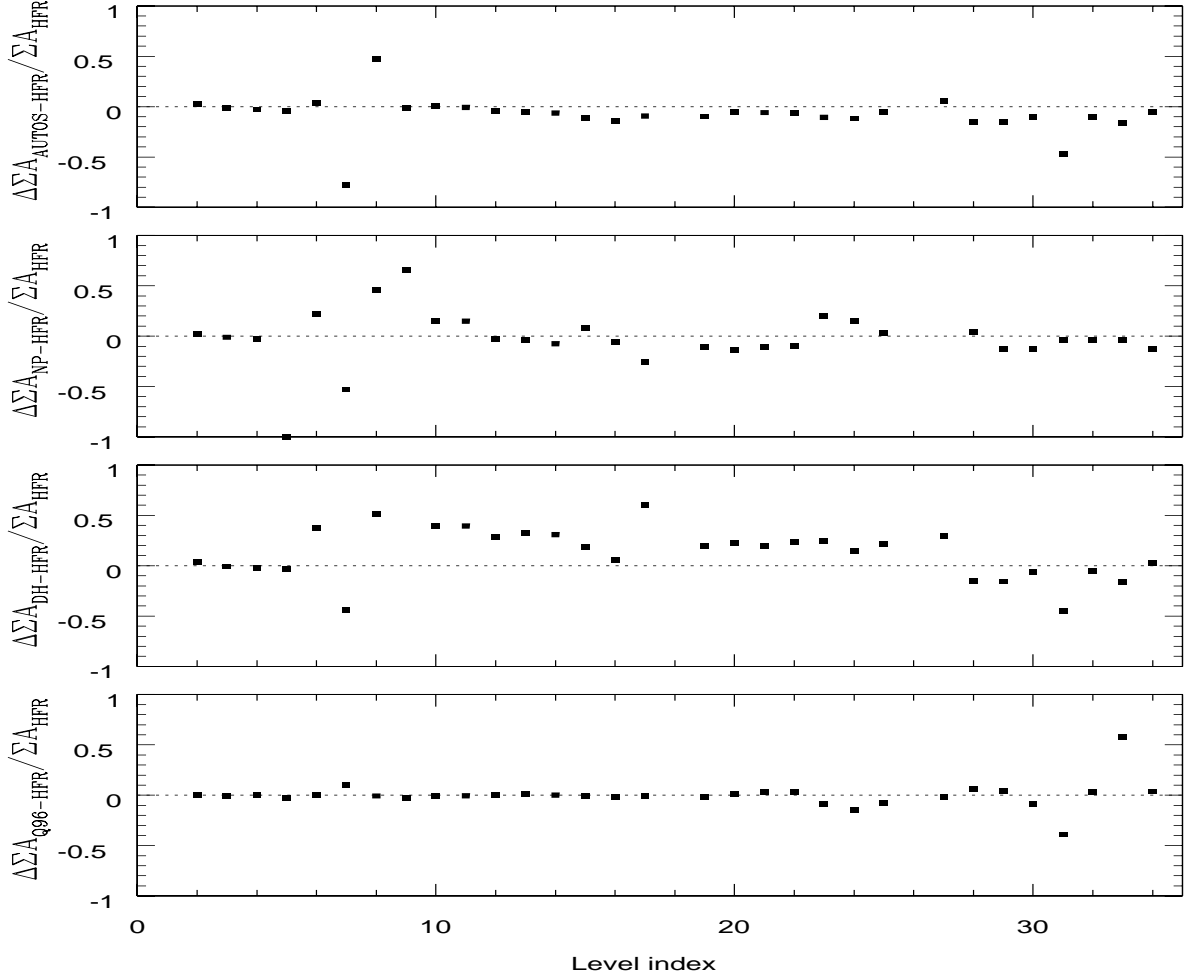}}
\caption{Comparison between present HFR results ($\sum A_{HFR}$) and those of Quinet (1996, $\sum A_{Q96}$), 
{\sc superstructure} calculations of Nahar and Pradhan (1996, $\sum A_{NP}$), and
Deb and Hibbert (2009). In each panel the y-axis gives the fractional 
differences with respect to the present HFR results, 
$\Delta\sum A_{X-HFR} = (\sum A_{X}-\sum A_{HFR}$). 
The x-axis of the plots corresponds to the level indexes, which are
assigned by strict increasing energy ordering. Levels 18 and 26 are missing from the figures because they belong to the 3d$^5$4s configuration.}
\label{hfrconverge} 
\end{figure}

In Fig.~\ref{hfrconverge} we compare the
results of the present HFR and AUTOSTRUCTURE calculations. We also
compare these values with those of Quinet (1996), \cite{naharpradhan}, and 
\cite{debhibbert}. 
The comparisons are made in terms of $\sum_{i>j}A_i$ which is of more practical interest than individual A-values. 
This is because $\sum_{i>j}A_i$ values
are weighted towards the strongest transitions, which dominate the de-excitation of levels as well as the observed spectra.
The comparison is for levels of the 3d$^6$ configuration, which are the only ones reported in previous publications.
In comparing the present results with those of \cite{naharpradhan} one finds 
large differences for levels of the $^3$P and $^3$H terms and a 
dispersion of 
20~-~30\% for the rest. 
There is significant scatter ($\sim$50\%) between the results of all other calculations and
those of \cite{debhibbert}, who used the CIV3 package. The only
A-values in agreement are those among levels of the a~$^5$D ground term 
(see Section 4). 
When comparing the present results with those of Quinet (1996), also from HFR, 
one finds excellent agreement for all but four levels. There is also good 
agreement between the new AUTOSTRUCTURE and HFR ($\sim 10 - 20\%$), which 
suggests that the calculations have reached sufficient convergence for the most part. 
The only two problematic levels here are a~$^3$H$_6$ and a~$^3$H$_5$
to which we have to assign a large uncertainty. 

\section{Collision Strengths}

Collision strengths for \ion{Fe}{3}
were computed by two different methods,  
RM+ICFT and DARC. 
For the RM+ICFT calculation we use the orbitals from our AUTOSTRUCTURE calculation, retaining CI from all 36 configurations. Yet, 
due to computational constrains the close coupling expansion only includes 136 LS terms (283 levels) from the
3d$^6$, 3d$^5$4s, and 3d$^5$4p configurations. 
The calculation explicitly includes 
partial waves from states with $L\le 12$ and multiplicity 1, 3, 5, and 7.
The final collision strengths are produced with an energy resolution of
$6\times 10^{-5}$ Ry. The computations were performed with the suite of parallel BP
$R$-matrix programs \citep{Mitnik01,Mitnik03,Ballance04}

The DARC calculation was based on target orbitals determined
by the Dirac-Hartree-Fock atomic structure package GRASP 
\citep{Dyall89, Parpia96} within an extended average level (EAL) approximation. 
 The CI expansion of the
target included the 8 configurations ${\rm 3s^23p^63d^6}$, ${\rm 3s^23p^43d^8}$,
${\rm 3p^63d^8}$, ${\rm 3s^23p^53d^7}$, 
${\rm 3s^23p^63d^54s}$,        
${\rm 3s^23p^63d^54p}$,${\rm 3s^23p^43d^74s}$ and ${\rm 3s^23p^43d^74p}$
for a total of 2468 levels. The accuracy of the target states could have been further improved by including
the configurations ${\rm 3s^23p^63d^54d}$ and ${\rm 3s^23p^43d^74d}$,
but this was found to make the $R$-matrix calculation prohibitively large.

The scattering calculation was performed with the set of
parallel DHF $R$-matrix programs \citep{Ballance06,Badnell04}, which uses 
modified subroutines based upon the serial version of the {\sc DARC} code of \citep{Norrington04} 
and portions of the RM 
programs.  The 
close coupling expansion of the target included the 322 levels arising from the configurations
${\rm 3s^23p^63d^6}$, ${\rm 3s^23p^63d^54s}$ and ${\rm 3s^23p^63d^54p}$.
Fortunately, the last two configurations of our CI expansion
 do not 
support energy levels below 7 Ryd. 
 All ${\rm J\Pi}$ partial waves from ${\rm J = 0}$ to ${\rm J = 29.5}$
were included in the calculation and contributions from the higher 
partial waves were estimated from a \cite{Burgess} top-up procedure. For partial waves  up to ${\rm 9.5}$, 
we employed 11 basis orbitals for each continuum-electron angular momentum.
This continuum
basis was sufficient to span electron energies up to 3.0 Ryd.

\begin{deluxetable}{cccc}   
\tabletypesize{\scriptsize}\label{upsilon} 
\tablecaption{Maxwellian averaged collision strengths at 10,000~K from the ground
level 3d$^6$~$^5$D$_4$ to excited levels of the 3d$^6$.}  
\tablehead{
\colhead{Upper level} &  \colhead{RM+ICFT} & \colhead{DARC} & \colhead{Zhang} } 
\startdata 
$^5$D$_3$       & 4.57E+0 & 2.54E+0 & 2.92E+0 \cr
$^5$D$_2$       & 1.94E+0 & 1.11E+0 & 1.24E+0 \cr
$^5$D$_1$       & 8.79E-0 & 5.33E-1 & 5.95E-1 \cr
$^5$D$_0$       & 2.51E-1 & 1.60E-1 & 1.80E-1 \cr
$^3$P2$_2$      & 7.14E-1 & 7.14E-1 & 5.80E-1 \cr
$^3$P2$_1$      & 1.84E-1 & 1.96E-1 & 1.65E-1 \cr
$^3$P2$_0$      & 3.83E-2 & 3.25E-2 & 2.13E-2 \cr
$^3$H$_6$       & 2.66E+0 & 1.21E+0 & 1.34E+0 \cr
$^3$H$_5$       & 1.10E+0 & 9.84E-1 & 4.89E-1 \cr
$^3$H$_4$       & 2.41E-1 & 5.33E-1 & 9.26E-2 \cr
$^3$F2$_4$      & 1.47E+0 & 4.54E-1 & 1.07E+0 \cr
$^3$F2$_3$      & 6.42E-1 & 1.91E-1 & 4.35E-1 \cr
$^3$F2$_2$      & 2.11E-1 & 1.73E-1 & 1.57E-1 \cr 
$^3$G$_5$       & 1.11E+0 & 1.36E+0 & 1.10E+0 \cr 
$^3$G$_4$       & 1.24E+0 & 1.11E+0 & 4.28E-1 \cr 
$^3$G$_3$       & 4.52E-1 & 4.21E-1 & 1.09E-1 \cr 
\enddata 
\end{deluxetable}  

\begin{figure}\resizebox{\hsize}{12cm}{\includegraphics{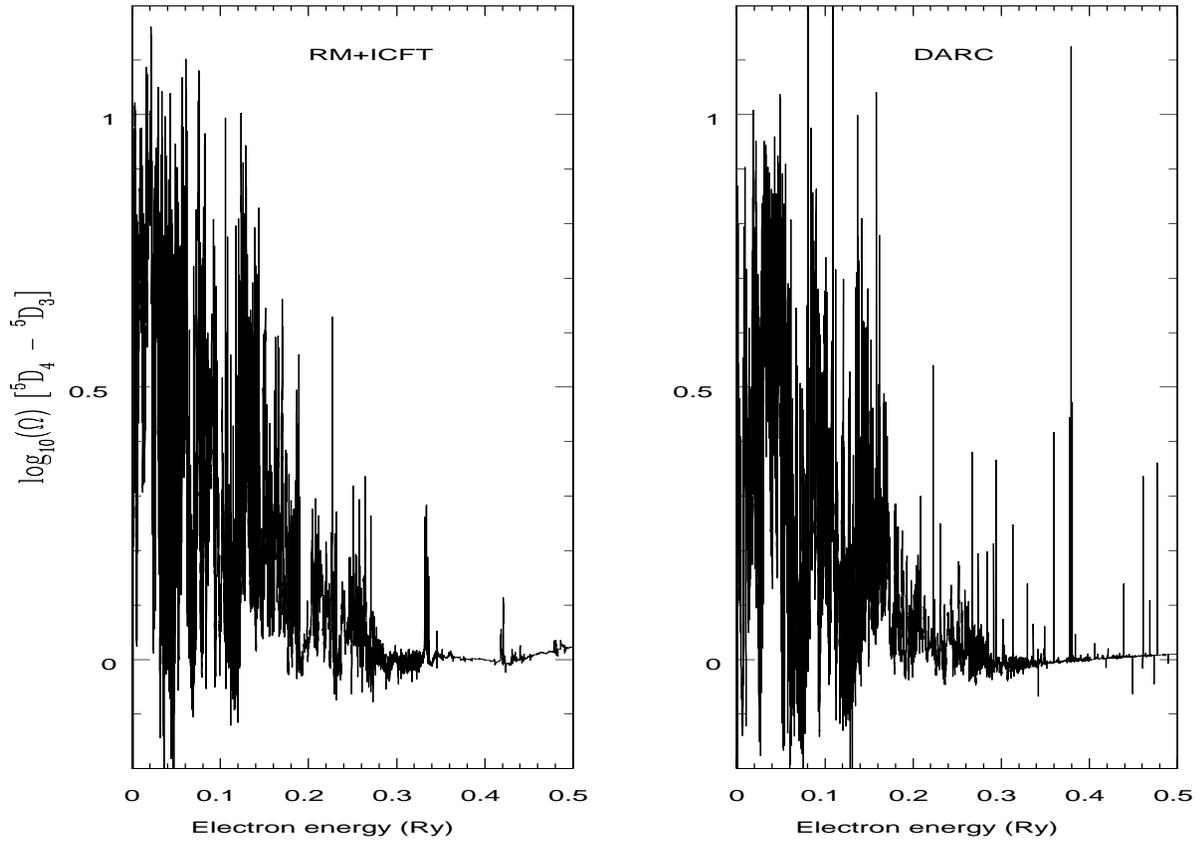}}
\caption{Collision strengths for the 3d$^6$~$^5$D$_4$~-~$^5$D$_3$ transition computed with RM+ICFT and DARC.
}
\label{omegas}
\end{figure}

We find that while RM+ICFT and DARC yield very similar background collision strengths, 
there are slight differences in the positions of the resonances near threshold.
This is illustrated in Figure~2, where one can see that a dense resonance structure is shifted towards the threshold in the RM+ICFT calculation with respect
to the DARC result. 
These resonances dominate the Maxwellian averaged collision strengths at temperatures near $10^4$~K, typical of photoionized plasmas.
For this reason the RM+ICFT Maxwellian averaged values are  
systematically higher than those from DARC by as much as $\sim 50\%$,
as shown in  Table~1 for a sample of transitions.
Also, there are large differences between the present results and those of \citep{zhang96} for many transitions.
The calculations of Zhang where carried out in LS-coupling  plus algebraic splitting of collision strengths to fine structure.

\section{Comparison with observed astronomical spectra}

We benchmarked the quality of the various data sets against spectra of the 
Orion nebula by \cite{mesadelgado}. This deep, high resolution spectrum shows 
33 optical and near infrared lines in a shocked gas region, "shock component", 
and 16 lines in a more diluted region, "nebular component". The electron 
densities, 
$n_e$, are diagnosed from lines of $[$\ion{O}{2}$]$, $[$\ion{S}{2}$]$, $[$\ion{Cl}{3} $]$,
and $[$\ion{Ar}{4}$]$ in both regions as $2890\pm 550$~cm$^{-3}$ and $17430\pm 2500$~cm$^{-3}$ respectively.
The electron temperature, $T_e$, in both regions is approximately 9000~K.

\begin{figure*}\resizebox{\hsize}{14cm}{\includegraphics{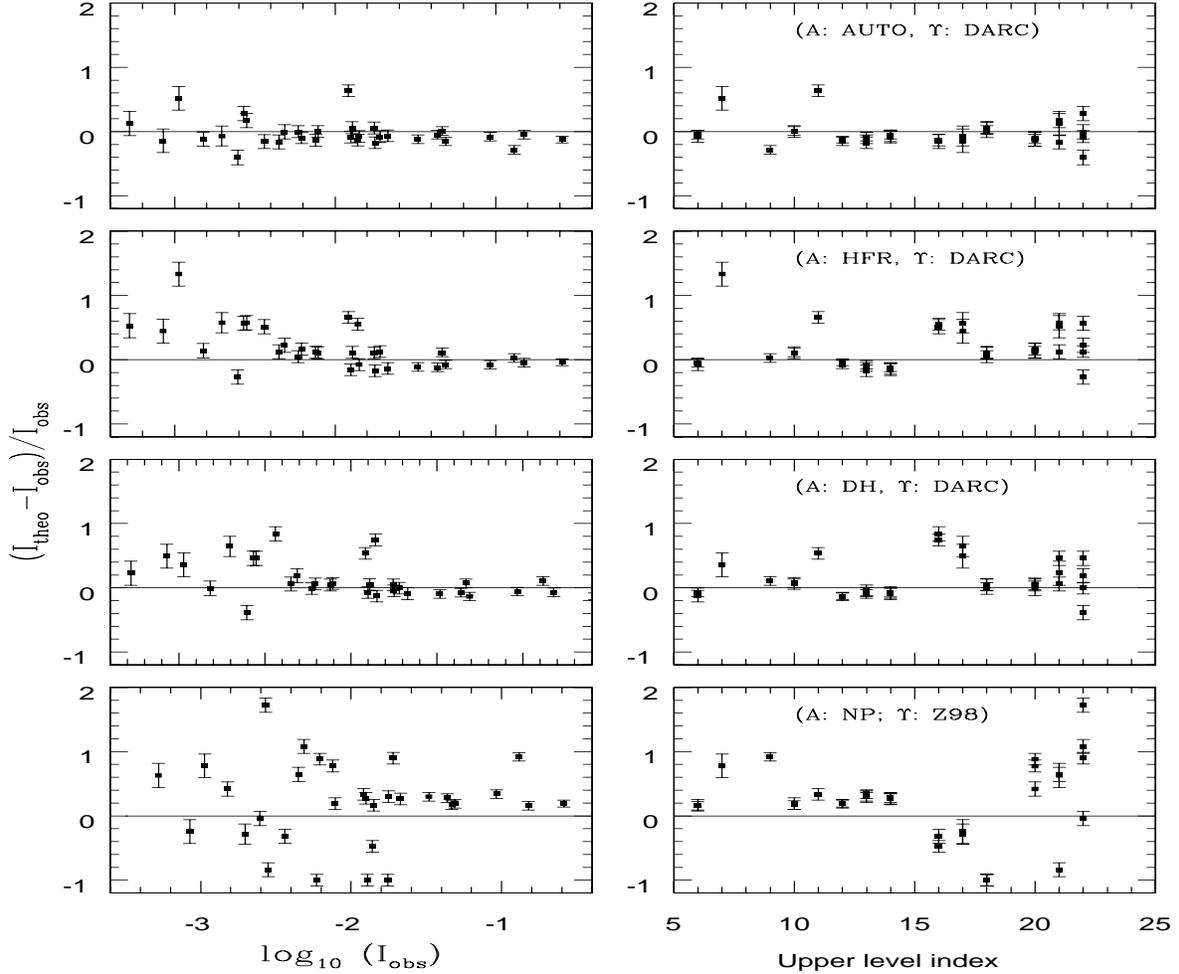}}
\caption{Comparison between observed and predicted line intensities for all 
lines measured in the "shock component" of the Orion Nebula.  
The left panels compare the lines intensities vs. the logarithm on the observed
intensity, and in the right panels the comparison are shown against the levels 
index of the upper level of the line. From top to bottom the first three rows
are for theoretical intensities using the present DARC collision strengths 
and A-values from AUTOSTRUCTURE, HFR, and Deb and Hibbert (HD) respectively,
while the lowest panels are based on theoretical intensities 
using collisional data from
\cite{zhang96} and A-values from \cite{naharpradhan}. The upper panel shows
the results from our present data.} 
\label{intensity}
\end{figure*}

We solve excitation equilibrium models for \ion{Fe}{3} using the various
sets of atomic data available adopting the values of
$n_e$ and $T_e$ given above. The fluxes of theoretical 
lines are normalized to the total flux predicted over all observed lines.
The fluxes of observed lines are also normalized to the sum of fluxes of all 
lines. 
Figure~\ref{intensity} shows the comparison between the theoretical and 
observed normalized intensities for the "shock component", which has the
richest spectrum. 
The comparison is in terms of the fractional difference between theoretical intensities and observations. The error bars in this figure are the statistical errors of the observations. 

We find that the previous model, that uses collision strengths from \cite{zhang96} and A-values from 
\cite{naharpradhan}, yield large scatter for all lines.
The new collisional data from DARC in combination with any of the newer sets
of A-values yield much better agreement with observations.
When comparing line intensities vs. the observed line intensity one sees that
the scatter increases towards the weakest lines, as expected from increasing 
theoretical and observational uncertainties for weaker lines. The comparisons 
of line intensities vs. the upper level index of the line 
allows one to see the effects of different sets on the population of each excited level. When two or 
more lines are observed from the same level the dispersion of these along the vertical direction indicates the uncertainties in the radiative
branching ratios for the decay of the level.
One can see that for levels 20, 21, and 22 ($^3$D$_{6,\ 5,\ 4}$) the Nahar and Pradhan A-values
yield considerable dispersion in branching ratios, unlike any of the 
other A-values datasets.
For levels 16 and 17 ($^3$G$_{4,\ 3}$) the combination of DARC
collision strengths and A-values from Deb and Hibbert and the present HFR
calculation yield line intensities overestimated by $\sim 60\%$, while 
the present AUTOSTRUCTURE A-values yield good agreement with observations.
For level 11 ($^3$P$_0$) all the newer A-values yield overestimated uncertainties by $\sim 50\%$, suggesting that the error could come from the collision
strengths, unless it is in the measured line intensity. The predicted population
of level 7 ($^3$H$_6$) seems too high in all models, but it is particularly 
problematic when using A-values from the present HFR calculation. 
The total radiative decay rate from this level as computed with HFR disagrees 
with the results of all other calculations. Thus, the observations seem
to provide evidence against the HFR rates for this level. 

When using the RM+ICFT collision strengths in the calculation of line
intensities, these are considerably worse than those from the data of Zhang (1996).
The large difference 
between the ICFT and DARC calculations is unexpected.
There was an expectation that given the extensive CI description of the target 
the ICFT method would compare more favorably with observations than
the DARC calculation, which uses  
only 7 configurations.
However, it appears that this relatively small DARC calculation
gives sufficiently accurate collision strengths. 
While the background collision strengths are very similar
among the ICFT and DARC calculations, this is not the case for near threshold resonances, which dominate the Maxwellian averaged values for forbidden transitions.
Representation of these resonances with RM+ICFT is fundamentally flawed 
because the transformation does not correct their positions by spin-orbit coupling.
Neither does the algebraic transformation used by Zhang. 

\section{Discussion and Conclusions}

We have carried out the most extensive and systematic calculation of radiative and collisional data for \ion{Fe}{3} done up to the present. We employed a 
multi-method approach that allowed  us to study the effects of different approximations and the advantages and shortcomings of each method.

In the calculation of A-values, we find convergence of the results in terms of 
methods and configuration expansions, the present ones being significantly larger than 
in previous work. There is general good agreement between the present 
AUTOSTRUCTURE and HFR results, despite significant discrepancies for a handful
of levels. The present results agree within $\sim 20\%$ with the result results
of Deb and Hibbert, but greater discrepancies are found with respect to 
those of Nahar and Pradhan (1998). It is difficult to pick the most accurate
of these calculations on the bases of the methods used alone (alhtough 
see Deb and Hibbert). Though, comparisons between 
predicted line intensities and measured lines in the spectra of the Orion
nebula seem to favor the present AUTOSTRUCTURE results over other 
calculations, and the A-values of Deb and Hibbert 
compare better with observations than those from HFR.

Benchmarking of collision strengths computed in LS coupling with transformation to intermediate coupling and those in jJ coupling sheds unexpected and important conclusions. 
Only calculations in jJ can accurately treat the near threshold resonances that dominate thermally averaged collision strengths at typical temperatures of photoionized nebulae. In the case of \ion{Fe}{3} the most important effect on resonance positions comes from spin-orbit coupling, and not so much from relativistic effects. In fact, algebraic splitting of LS collision strengths is practically as accurate as ICFT in terms of the background cross sections, but neither one can treat the near threshold resonances. Moreover, we see that the results of the present very large RM+ICFT calculations are somewhat worse than those of \citep{zhang96}. This is because the higher orbitals in the present CI expansion lead to a larger R-matrix box and this yields even poorer treatment of near threshold resonances and over a more extended energy region.

At this point one may ask whether Breit Pauli R-matrix (BPRM) calculations, i.e. in jK 
coupling, of
collision strengths could be an alternative to the use of DARC. The answer is
probably yes, although most BPRM packages available do not include two-body 
relativistic terms, but these should be unimportant here.
However, the main difference between the two approaches will be on the
quality of the atomic orbitals used. DARC uses orbitals from GRASP, which is
fully relativistic, thus optimizes on the energy levels and yields two
components to the radial functions of each orbital. This is ideal for representing strongly mixed levels. By contrast, AUTOSTRUCTURE orbitals to use in
BPRM calculations are optimized on LS terms only, thus greater CI 
expansions are often needed to reach the same degree of accuracy 
given by GRASP. 

The high accuracy of the present \ion{Fe}{3} spectral model, that
uses collisional data from DARC and A-values from AUTOSTRURE,
is demonstrated by 
comparison with observed optical and near infrared spectra of the Orion nebulae.
In absence of experimental determinations of atomic parameters, comparison 
with observed spectra is the best way to test the accuracy of the atomic
models. The present comparisons demonstrate that the new atomic data
is considerably better than previous results. 
Moreover, this study opens the door for the computation of long awaited accurate models for lowly ionized iron-peak species.

\begin{acknowledgements}
We acknowledge financial support from grants
from the NASA Astronomy and Physics Research and Analysis Program (award NNX09AB99G),
the Space Telescope Science Institute (project GO-11745), US
Department of Energy, and the 
Belgian F.R.S.-FNRSs, and from the Belgian F.R.S.-FNRS from which
PQ is Senior Research Associate. 
Much of the computations were carried out at the 
National Energy Research
      Scientific Computing Center in Oakland, California and at the
      National Center for Computational Sciences at Oak Ridge under the 
      Teragrid program.
\end{acknowledgements}



\end{document}